\documentclass{aastex631}

\usepackage{ulem}
\usepackage{soul}

\usepackage{amsmath}
\hypersetup{hypertex=true,
            colorlinks=true,
            linkcolor=blue,
            anchorcolor=blue,
            citecolor=blue}
\usepackage{hyperref}
\begin{document}

\title{Results of 15-Year Pulsar Timing of PSR J0007+7303 with Fermi-LAT}

\author{Zhi-xiang Yu}
\affiliation{School of Physics and Electronic Science, Guizhou Normal University, Guiyang 550025, China}

\author[0000-0002-2060-5539]{Shi-jun Dang} 
\email{dangsj@gznu.edu.cn(S.J.Dang)}
\affiliation{School of Physics and Electronic Science, Guizhou Normal University, Guiyang 550025, China}

\author{Wei-hua Wang}
\email{wang-wh@wzu.edu.cn (W. H. Wang)}
\affiliation{College of Mathematics and Physics, Wenzhou University, Wenzhou 325035, China}
   
\author[0000-0003-3321-9458]{Hui-hui Wang}
\affiliation{School of Physics and Engineering, Henan University of Science and Technology, Luoyang 471023, China}

\author{Lin, Li}
\affiliation{School of Physics Science and Technology, Xinjiang University, Urumqi, Xinjiang 830011, China}

\author[0009-0009-8247-3576]{Wei, Li}
\affiliation{Xinjiang Astronomical Observatory, Chinese Academy of Sciences, 150 Science 1-Street, Urumqi, Xinjiang 830011, China}
\affiliation{University of Chinese Academy of Sciences, Chinese Academy of Sciences, Beijing 100049, China}

\author{Jian-ping Yuan}
\affiliation{Xinjiang Astronomical Observatory, Chinese Academy of Sciences, 150 Science 1-Street, Urumqi, Xinjiang 830011, China}

\author[0000-0002-9786-8548]{Fei-fei Kou}
\affiliation{Xinjiang Astronomical Observatory, Chinese Academy of Sciences, 150 Science 1-Street, Urumqi, Xinjiang 830011, China}
\affiliation{Xinjiang Key Laboratory of Radio Astrophysics, 150 Science1-Street, Urumqi 830011, P. R. China}

\author{Jun-tao Bai}
\affiliation{Xinjiang Astronomical Observatory, Chinese Academy of Sciences, 150 Science 1-Street, Urumqi, Xinjiang 830011, China}

\author{Mingyu Ge}
\affiliation{Key Laboratory of Particle Astrophysics, Institute of High Energy Physics, Chinese Academy of Sciences, Beijing 100049, China}

\author{Xia Zhou}
\affiliation{Xinjiang Astronomical Observatory, Chinese Academy of Sciences, 150 Science 1-Street, Urumqi, Xinjiang 830011, China}
\affiliation{Key Laboratory of Radio Astronomy and Technology (Chinese Academy of Sciences), A20 Datun Road, Chaoyang District, Beijing, 100101, P. R. China}
\affiliation{Xinjiang Key Laboratory of Radio Astrophysics, 150 Science1-Street, Urumqi 830011, P. R. China}

\author{Lun-hua Shang}
\email{lhshang@gznu.edu.cn(L.H.Shang)}
\affiliation{School of Physics and Electronic Science, Guizhou Normal University, Guiyang 550025, China}

\author{Zu-rong Zhou}
\affiliation{National Time Service Center, Chinese Academy of Sciences, Xi’an 710600, China}

\author{Yu-bin Wang}
\affiliation{School of Physics and Electronic Engineering, Sichuan University of Science \& Engineering, Yibin 644000, China}

\author{Yan-qing Cai}
\affiliation{University of Chinese Academy of Sciences, Chinese Academy of Sciences, Beijing 100049, China}
\affiliation{National Astronomical Observatories, Chinese Academy of Sciences, Beijing 100101, China}

\author{Ru-shuang Zhao}
\affiliation{School of Physics and Electronic Science, Guizhou Normal University, Guiyang 550025, China}

\author{{Qing}-ying Li}
\affiliation{School of Physics and Electronic Science, Guizhou Normal University, Guiyang 550025, China}

\author{Xiang-dong Zeng}
\affiliation{School of Physics and Electronic Science, Guizhou Normal University, Guiyang 550025, China}

\author[0000-0003-2991-7421]{Na Wang}
\affiliation{Xinjiang Astronomical Observatory, Chinese Academy of Sciences, 150 Science 1-Street, Urumqi, Xinjiang 830011, China}
\affiliation{Key Laboratory of Radio Astronomy and Technology (Chinese Academy of Sciences), A20 Datun Road, Chaoyang District, Beijing, 100101, P. R. China}
\affiliation{Xinjiang Key Laboratory of Radio Astrophysics, 150 Science1-Street, Urumqi 830011, P. R. China}

\begin{abstract}
The study of pulsar glitches provides a unique window into the internal structure and dynamic processes of neutron stars. PSR J0007+7303, a very bright gamma-ray pulsar, is the first pulsar discovered by the Fermi-LAT telescope.  In this paper, we present the 15 years of timing results of this pulsar using the Fermi-LAT data. We identified nine glitches, five of which are newly discovered. Among these, two are small glitches, occurring between the three previously reported ones, while the other four are large glitches. The glitches exhibit fractional frequency changes ranging from $15 \times 10^{-9}$ to $1238 \times 10^{-9}$, with intervals of approximately 1–2 years between events. Uniquely, this pulsar shows no exponential recovery behavior following any glitch, setting it apart from most glitching pulsars. Furthermore, no significant changes were observed in the gamma-ray pulse profile, flux, or phase-averaged spectra before and after glitches, indicating the stability of the pulsar’s emission properties despite internal changes. A parametric analysis of the glitches yielded a fractional moment of inertia of the crustal superfluid involved in glitches as $1.06\%$, which matches extremely well with previous statistic work if the non-dissipative entrainment effect is not considered and strongly supports the internal origin of these glitches. These results highlight the distinct glitch behavior of PSR J0007+7303 and offer valuable insights into the crust-superfluid interaction in neutron stars. The physical origin of no exponential recovery is also discussed.

\end{abstract}


\section{Introduction} \label{sec:intro}
Pulsars are a special neutron star with high-speed rotation and a strong magnetic field \citep{Reichley1969}.
Typically, the rotation of a pulsar slows down gradually due to the loss of rotational energy \citep{Alpar2006,Taylor1979}, it can be predicted using a simple timing model with a Taylor expansion of the rotation period \citep{Edwards2006}. The pulsar timing technique allows precise measurements of the pulsar's rotational and astrometric parameters, such as the rotation frequency, spin-down rate, position, proper motion, and so on. This technique mainly involves long-term and frequent timing monitoring of pulsars to obtain the time of arrival(ToA) of the pulsar, and then using a pulsar timing model to fit the ToAs, thereby obtaining the rotation parameters and astrometric parameters of the pulsar.  The pulsar timing technique is important for the study of the long-term spin evolution of pulsar,
and it has also been applied to detect the internal state of neutron stars, interstellar medium, pulsar navigation \citep{Deng_2013}, detection of gravitational waves using pulsar timing arrays \citep{Kramer2009}, and test of general relativity \citep{Hobbs2004,Stairs2003}.

Timing noise and glitches are two main phenomena that cause irregularities in pulsar timing solutions. Timing noise, a low-frequency phenomenon that has not yet been precisely modeled, often exhibits quasi-periodic characteristics and can be found in the observational residuals of pulsars after all known and predictable spin-down effects have been eliminated~\citep{Cordes1985,Arzoumanian1994,Hobbs2010}.
The glitch of pulsars is manifested as a sudden spin-up of the rotation frequency ($\nu$) and usually a step increase in spin down rate ($\dot\nu$) of pulsars, followed by a gradual recovery, with a recovery timescale usually ranging from a few days to hundreds of days~\citep{Yuan_2010,Dang2020,Zhou2022,Li2023}.
There exists two mainstream models accounting for pulsar glitches, namely, the starquake model and the superfluid vortex model~\citep{Haskell2015}.
Starquake occurs due to adjustment of neutron star shape and crustal stress release during the long term spin down~\citep{1969Natur.223..597R,1971AnPhy..66..816B}, when starquake happens, the moment of inertia of neutron stars decreases suddenly, while the spin frequency increases abruptly due to angular momentum conservation.
For the superfluid vortex model, the crustal superfluid component of the neutron star rotates faster than the crystalline crust, the angular momentum exchange between these two components will result in the glitch~\citep{1969Natur.224..673B,1975Natur.256...25A}.
The magnitude of the glitch is measured by the change in frequency ($\Delta \nu$) or the fractional change in frequency ($\Delta\nu/\nu$). 
Generally, the fractional increase in rotational frequency lies between $\Delta \nu / \nu \sim 10^{-9}-10^{-6}$. 
The distribution of $\Delta \nu$ from all known glitches exhibits bi-modality~\citep{2017A&A...608A.131F}.
There are strong correlations between glitches, characteristic age, and the first-order derivative of the rotation frequency. 
The glitch is more easily observed in younger pulsars and decreases with age until it disappears~\citep{2011MNRAS.414.1679E}. 
Moreover, the large glitch is more likely to occur in young pulsars~\citep{Yuan_2010,Basu2022}. 
Studying the glitch of pulsars can obtain a lot of information about the interior of neutron stars~\citep{Link1999,Andersson2003}. 
The recent studies found that both the timing noise and glitches in some pulsars are correlated with their emission variations~\citep{2010Sci...329..408L,Takata2020,2022MNRAS.513.5861S}.
Studying the connection between the pulsar glitch phenomenon and pulse radiation can further our understanding of neutron stars~\citep{Weltevrede2011,Kou2018,Takata2020,Wang2024}.

Pulsars are one of the major sources of gamma-rays in the Milky Way galaxy. Long-time observations of pulsars in the radio band have revealed a large number of glitches, but some pulsars have detected radiation only in the gamma-ray band without the radio part. The Fermi-LAT telescope provided the convenience for the timing study of these radio-quiet gamma-ray pulsars. The study of the glitch phenomenon of gamma-ray pulsars can increase the glitching pulsar samples and contribute to the scientific study of high-energy astrophysics \citep{Ray_2011}.

PSR J0007+7303 is a bright gamma-ray pulsar and and located in a supernova remnant with a characteristic age of about $\sim$13.9\ kyr, located at 1.4 ± 0.3\ kpc, consistent with CTA 1(5 $\sim$ 15\ kyr) \citep{Pineault1993}. Similar to Geminga and PSR J1836+5925, PSR J0007+7303 is a radio-quiet gamma-ray pulsar. Previous study shows that three glitches have been detected in this pulsar, and no exponential recovery was found after its glitches\citep{2016ApJ...831...19L}. In this paper, we perform a timing analysis of the gamma-ray pulsar PSR J0007+7303, using 15 years (from August 4, 2008 to October 24, 2023) of observations from the Fermi-LAT telescope.
In section~\ref{sec:Obs}, we describe the observations and the data processing. In section~\ref{sec:Results}, we report the results of the timing analysis of this pulsar. The sections~\ref{sec:Discussion} and~\ref{sec:Summary} are discussion and summary of our results, respectively.

\section{Observation and Data analysis }
\label{sec:Obs}
The Fermi space station was launched in 2008 and consists of two telescopes, one is the GBM (Gamma-Ray Bursting Machine) and the other is the LAT (Large Area Telescope). The energy band of the LAT ranges from $\sim$20\ MeV to \textgreater 300\ GeV, and the energy resolution is \textless 15\% of the energy \textgreater 100\ MeV. Fermi-LAT has a very large field of view, and it completes a survey of the whole sky in about three hours, with an effective area of \textgreater 8000  $\rm cm^{2}$\citep{Abdo2010}. The LAT telescope is a very powerful tool for the observation of  gamma-ray pulsars, and it has discovered about 340 gamma-ray pulsars in 15 years \citep{Ray_2011,2023ApJ...958..191S}.

In this work, we use the LAT PASS 8 data (P8R3, from August 4, 2008, to October 24, 2023) to perform timing analysis of PSR J0007+7307, which is analyzed with the Fermi Science Tools(v11r5p3)\footnote{https://fermi.gsfc.nasa.gov/ssc/data/analysis/scitools/pulsar$\_$analysis$\_$tutorial.html}.
We select the photon events in an energy range of 150$-$10,000 Mev, with angular distance less than 3$^\circ$, which maximized the H-test value, and set at a zenith angle of less than 90$^{\circ}$ to minimize gamma-ray secondary radiation from the Earth's atmosphere in the data\citep{Ge_2020}. The above filtering is accomplished by the ``gtselect" task, and only photon events included in these ranges were included in the timing analysis. Normally we would like to measure time of arrival (ToAs) at a fixed position on Earth, but the Fermi space station is in periodic orbital motion around the Earth. Therefore, we correct the time to DE405 ephemeris using ``gtbary" \citep{Standish1998,10.1093/mnras/stac026}, the time system for photon events remains Terrestrial Time (TT). Next, assign phases to the photons according to the local ephemeris using the fermi-plugin of \textsc{Tempo2} \citep{Ray_2011,Edwards2006}.
We use the high signal-to-noise ratio template to correlate with the unbinned centered data to obtain the ToA according to each phase. Here, each ToA is accumulated for 30 days of photons in order to obtain sufficient accuracy in the arrival time of ToA.
Subsequently, we used the \textsc{Tempo2} software package\citep{2006MNRAS.369..655H} to perform a simple fitting of the ToAs to obtain the rotation parameters and the timing residuals. The timing fitting model we used is a Taylor expansion of the rotation phase $\phi(t)$, as follows:
\begin{equation}
    \phi(t) = \phi_0 + \nu_0(t - t_0) + \frac{1}{2}\dot{\nu}_0(t - t_0)^2 + \frac{1}{6}\ddot{\nu}_0(t - t_0)^3 + \cdots 
\label{Eq:Tming_model}
\end{equation}
where $\nu$, $\dot{\nu}$, $\ddot{\nu}$ represent the rotation frequency and its first and second-order derivatives, respectively. $\phi(t_{\rm 0})$ is the pulse phase at $t_{\rm 0}$, and it can be assumed that $\phi(t_{\rm 0}) = 0$ at $t = t_{\rm 0}$.

\begin{figure}
	\includegraphics[width=0.7\columnwidth,angle=-90]{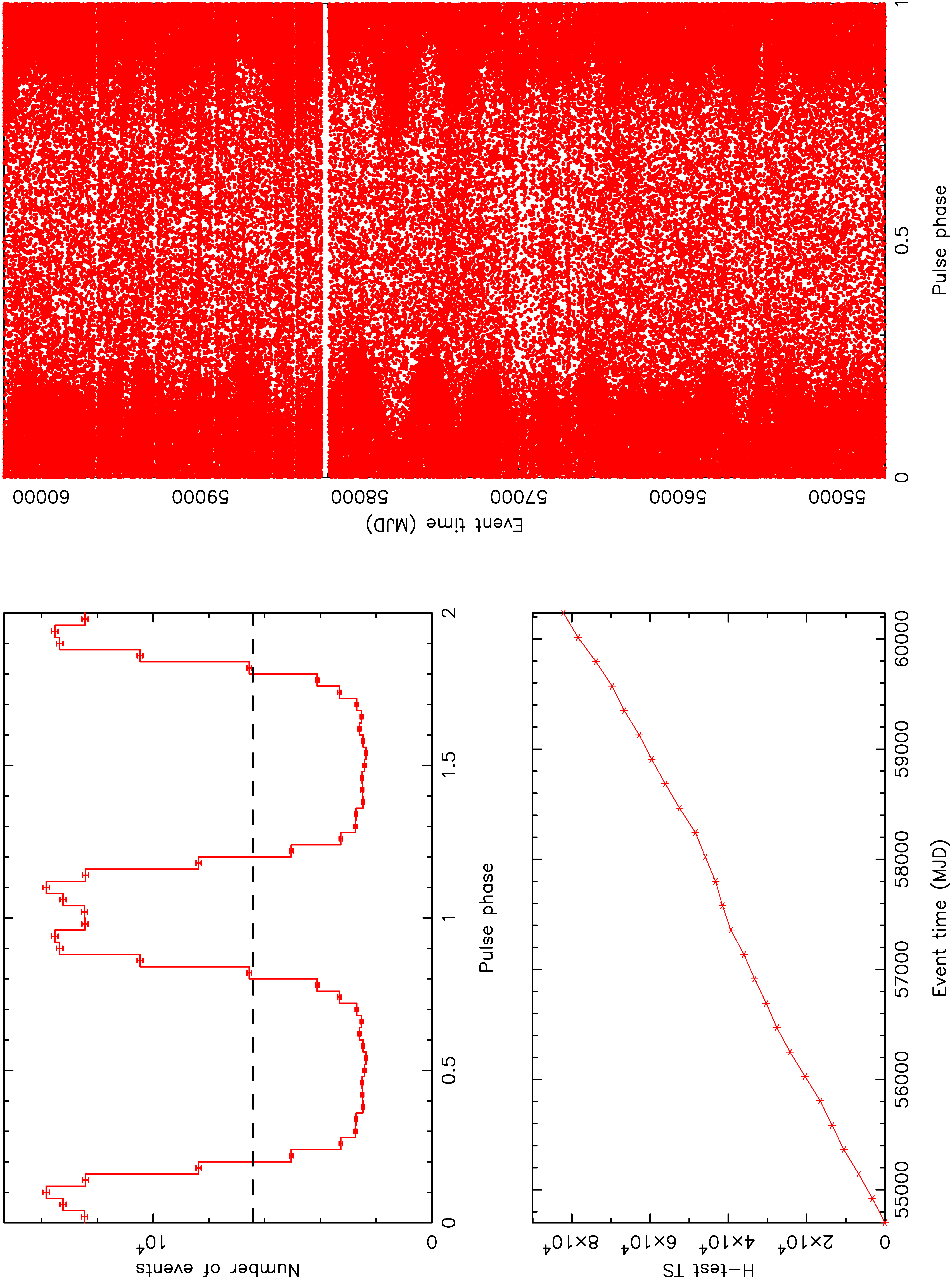}
    \caption{Phase alignment results of photons for the full data span of PSR J0007+7303. Top-left
panel: phase histogram of the analyzed gamma-ray data. Two full rotations are shown
for clarity. Bottom-left panel: H-test significance (TS) as a function of time. Right panel: pulse phase for each gamma-ray event versus time.}
    \label{total_profile}
\end{figure}

If the fitting process performs well, the timing residuals shows a distribution around zero. By analyzing the timing residuals we can determine whether glitching occurs, and we can initially determine whether glitching occurs by looking at the H-test plot. Further confirmation of whether glitching occurs is obtained by analyzing the timing residuals. When the timing residuals behave discontinuously after a certain time, a glitch has occurred in that period~\citep{Lin2021}. The change in pulse phase caused by glitches can usually be fitted using the following model~\citep {Edwards2006}:
\begin{equation}
\label{equ:1} 
\phi_{g}=\Delta\phi+\Delta\nu_{p}(t-t_{g})+\frac{1}{2}\Delta\dot\nu_{p}(t-t_{g})^{2}+[1-e^{-(t-t_{g})/\tau_{d}}]\Delta\nu_{d}\tau_{d},
\end{equation}
where $t_{g}$ is the time of glitch, $\Delta\phi$ is the increment of pulse phase in the data before and after the glitch.
Due to the frequent glitches and strong timing noise of this pulsar, the ephemeris is not accurate enough, resulting in the inability to align the photon phases, which in turn leads to large ToA uncertainty. Therefore, we repeat the above steps until the Root Mean Square (RMS) of the timing residuals no longer reduces, so as to align the photon phases as much as possible. Figure \ref{total_profile} shows the phase–time diagram of phase alignment for PSR J0007+7303 obtained using the timing solution over the full data span, along with the corresponding integrated pulse profile and the evolution of the H-test Test Significance (TS) value over time. The TS value increases linearly with time, indicating the validity of the adopted timing ephemeris. Finally, we utilized the easyFermi software\citep{2022A&C....4000609D} to analyze the changes in the phase averaged spectrum before and after each glitch.
\section{Result} \label{sec:Results}

\subsection{Timing solutions and glitch parameters}

\begin{figure*}

\centering
\includegraphics[width=0.7\columnwidth,angle=-90]{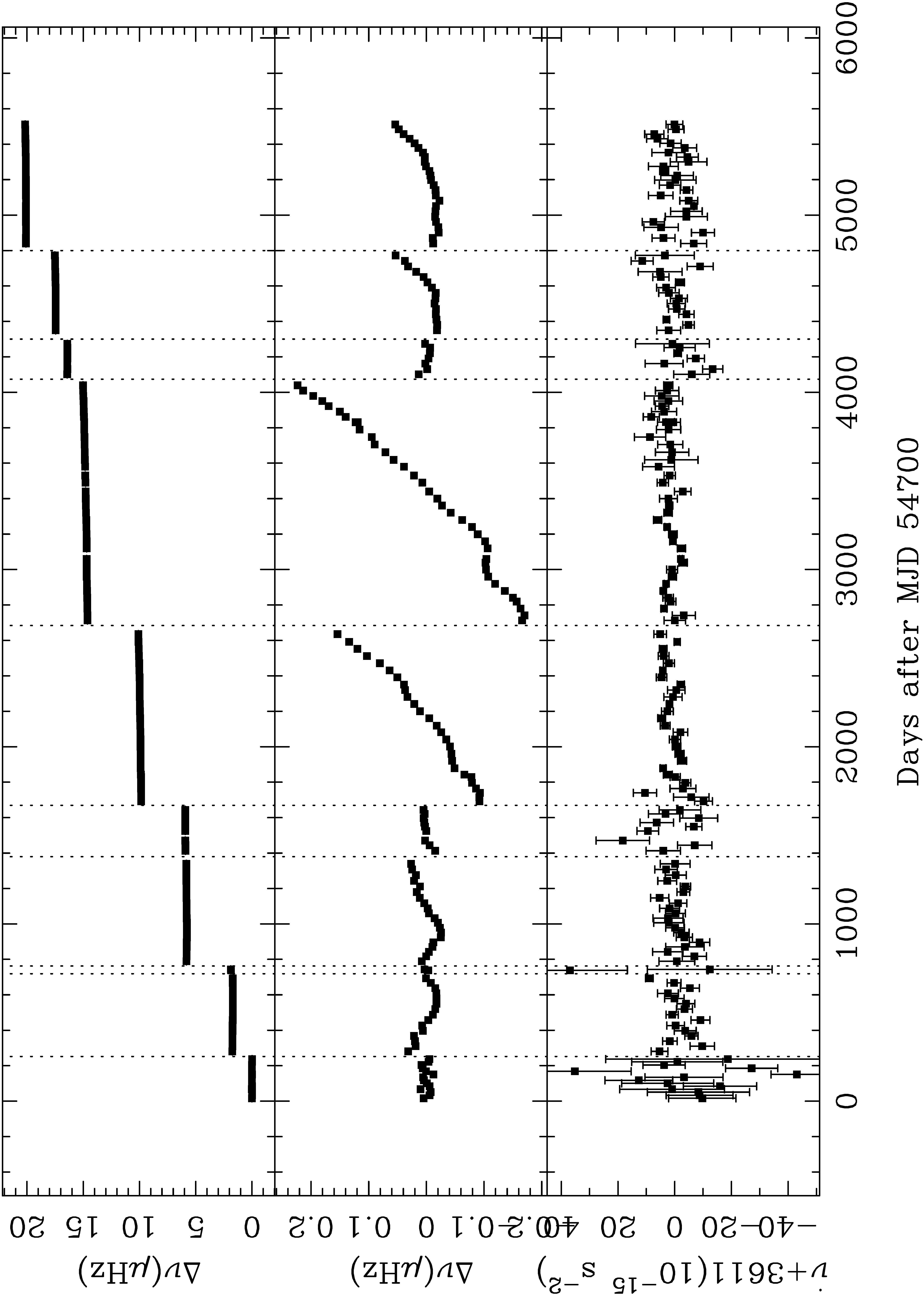}
\begin{center}
\caption {Glitches in PSR J0007+7303. The top panel shows the frequency residual ($\Delta \nu$), the middle panel represents the evolution of frequency residual ($\Delta \nu$) after removing mean frequency of post-glitch, the bottom panel is the derivative of the spin frequency, with the mean value removed. The vertical dashed lines represent the glitches epoch.}
\label{all_glitch}
\end{center}\vspace{-0.4cm}
\end{figure*}

\begin{figure}
	\includegraphics[width=0.5\columnwidth ,angle=0]{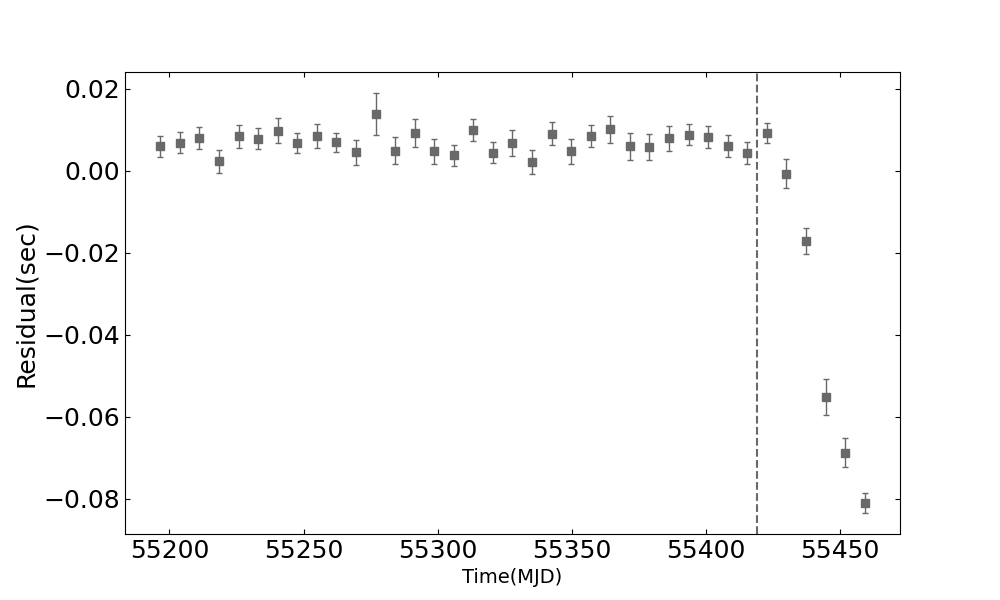}
 \includegraphics[width=0.5\columnwidth,angle=0]{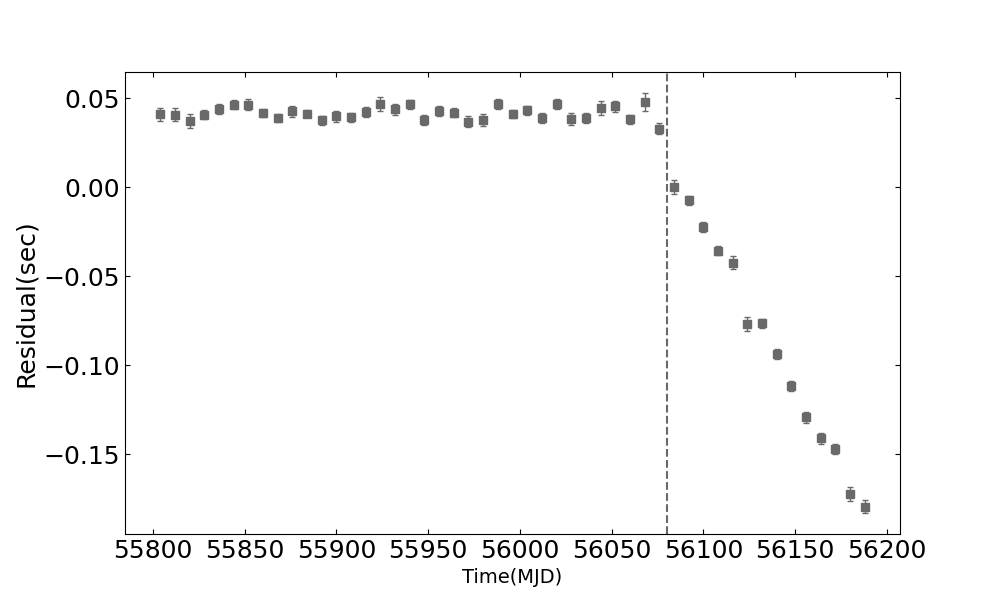}
    \caption{The timing residuals relative to pre-glitch models of PSR J0007+7303 around the second(left) and fourth(right) glitches. The vertical dashed lines stand for the glitch epoch.}
    
    \label {the second glitch}
\end{figure}

By analyzing the timing data of PSR J0007+7303 for a 15-year data span (From August 2008 to October 2023), we detected nine glitch events, four of which have been reported by The Third Gamma-ray Pulsar Catalog (3PC,\cite{2023ApJ...958..191S}),  while the remaining five are new glitches discovered in this work. Additionally, we have detected two minor glitches within the glitch intervals reported by \citet{2016ApJ...831...19L}.
To study the variations of frequency $\nu$ and its first derivative $\dot\nu$ during our data span, we derived the $\nu$ and $\dot\nu$ by fitting short data sections, and each section containing about 60 ToAs and repeating for 30 ToAs. Figure \ref{all_glitch} displays the evolution of $\nu$ and $\dot\nu$. The top panel of Figure~\ref{all_glitch} shows the frequency residual ($\Delta \nu$), the middle panel represents the evolution of $\Delta \nu$ after removing the mean frequency of post-glitch, the bottom panel is the evolution of the spin-down rate after removing the mean value. The vertical coordinate is the magnitude of the change of the rotation frequency, the horizontal coordinate is the time (MJD), and the starting point at 0 represents MJD 54700, black data points represent the residuals of the frequency, the vertical dashed lines indicate glitch epoch and an obvious discontinuity can be observed in frequency evolution around glitch epoch. Glitches divide the entire time span into ten segments, the interval between the second and the third glitch is the shortest in this work, which is only about 44 days. The frequency variations of the second glitch and the fourth glitch are not obvious because their sizes are much smaller than the other glitch events. The lower part of Figure~\ref{all_glitch} demonstrates the evolution of the $\dot\nu$ without a significant exponential recovery. The pre-fit timing residuals in Figure~\ref{the second glitch} show a significant linear deflection at MJD 55419(2) and MJD 56080(2), which is consistent with the typical pattern observed in small glitches. In this pulsar, we have not detected significant exponential recovery features after glitches. 

\begin{table*}
\centering
\tiny
\caption{Pre- and post-glitch timing solutions for PSR J0007$+$7303.} 
\label{pre and post glitch solutions}
\begin{tabular}{cccccccccccc}
  \hline
    \hline        
RA      &DECL   &POSEPOCH  & PEPOCH  & $\nu$  & $\dot{\nu}$ & $\ddot{\nu}$  & $N_{\rm ToA}$    & MJD Range   & RMS  & n  \\
 (h:m:s) &(d:m:s) &(MJD)     & (MJD)  & (Hz)   & ($10^{-14}$ s$^{-2}$)   &  ($10^{-24}$ s$^{-3}$) &  & (MJD) & ($\mu$s) &  \\
    \hline 
&  &    &54825  &3.1658670273(7)  &$-$361.20(9)   &$-$7(49)  &31    &54704 -- 54952  &2861.211 &$-$1.7(12)  \\     
&  &    &55207  &3.1657495093(19)  &$-$361.50(5)   &409(63)  &29    &54960 -- 55416  &2470.751 &99(15)   \\ 
&  &    &55419  &3.1656833626(26)  &$-$360.75(6)   &410(62)  &6     &55419 -- 55463  &2693.258 &100(15)  \\ 
&  &    &55631  &3.1656212657(5)  &$-$361.21(1)   &75(8)  &76     &55464 -- 56073  &10621.989  & 18(2) \\    
00:07:01.58(8) &$+$73:03:07.3(4) &56749   &56362  &3.1653932556(38)  &$-$360.99(11)   &286(145)  &23     &56081 -- 56362  &2441.129  &70(35) \\   
&  &    &56749  &3.1652764512(2)  &$-$360.971(2)   &$-$10(3)      &77   &56368 -- 57379  &4312.152  &  $-$2.4(8)\\  
&  &    &57999  &3.1648914272(4)   &$-$360.928(2)    &43(2)    &136   &57387 -- 58492  &17525.882 & 10.3(4)   \\ 
&  &    &58884  &3.1646165881(6)   &$-$361.31(2)   &98(67)    &26   &58765 -- 58997  &2013.811    &   24(16)\\  
&  &    &59249  &3.1645037137(3)   &$-$361.089(4)   &186(4)    &58    &59005 -- 59486  &4907.776   &  45(1)\\  
&  &    &59960  &3.1642844576(5)   &$-$361.099(6)    &59(6)      &88    &59502 -- 60230 &13223.704  &   14(2)\\        
\hline
\end{tabular}
\end{table*}

\begin{table*}
\large
\caption{The fitted timing solutions and glitch parameters for PSR J0007$+$7303. The timing solution uncertainties in Table are expressed as 1$\sigma$ values obtained through \texttt{TEMPO2}.} 
    \begin{tabular}{clllllcc}
    \hline
    Gl.No. & Glitch Epoch & $\Delta\nu$ & $\Delta\dot\nu$ & $\Delta\nu/\nu$ & $\Delta\dot\nu/\dot\nu$  \\
    	& (MJD) & (10$^{-6} {~\rm Hz}$) & (10$^{-15} {~\rm s^{-2}}$) & (10$^{-9}$) & (10$^{-3}$)	\\
    \hline 
1 & 54952(2)   & 1.752(3)   & $-$2.2(23)     & 553.4(1)    & 0.6(6)      \\
2 & 55419(2)   & 0.083(15)  & $+$3.2(32)    & 26(19)       & $-$0.9(9)    \\
3 & 55463(2)   & 3.892(67)  & $-$3.2(32)   & 1230(21)     & 8.9(90)      \\
4 & 56080(2)   & 0.068(3)   & $-$2.03(23)    & 21.5(9)       & 0.6(6)      \\
5 & 56369(2)   & 3.92(3)   & $-$1.7(23)    & 1239(10)     & 0.5(6)       \\
6 & 57381(3)   & 3.37(4)   & $-$3.7(2)     & 1065(2)     & 1.0(3)       \\
7 & 58771(3)   & 1.377(3)   & $-$7.5(9)     & 435.1(30)   & 2.1(3)        \\
8 & 59000(3)   & 1.039(10)   & $+$1.5(9)     & 328.3(32)    &$-$0.4(3)     \\
9 & 59498(2)   & 2.59(3)   & $-$2.7(2)    & 819(11)      & 0.7(41)        \\
    \hline
    \end{tabular}
\label{glitch solutions}
\end{table*}

Table~\ref{pre and post glitch solutions} shows the detailed parameters before and after each glitch, the uncertainties of the reported parameters are 1 $\sigma$ derived from \textsc{TEMPO2}. Table \ref{glitch solutions} gives the detailed parameters of each glitch. The first two columns show the glitch numbers and glitch epochs. The glitch epoch is obtained by the external derivation method, in which the midpoint between the last observed epoch (T1) before the glitch and the first observed epoch after the glitch is defined as the glitch epoch. We define the $1\sigma$ uncertainty of the glitch epoch as (T2-T1)/4 \citep{2011MNRAS.414.1679E,Dang2020}. The last four columns report the change in the rotational frequency and spin-down rate, as well as their relative changes ($\Delta \nu/\nu$ and $\Delta \dot\nu /\dot\nu$).

We update the glitch parameters of the three previous glitches and summarize detailed parameters of the nine glitch events in Table \ref{glitch solutions}. We derive the parameters by fitting ToAs to equation \ref{equ:1} using {\sc Tempo2}. The associated uncertainties are calculated utilizing the error transfer equation. The first, third, and fifth glitches have been reported previously, we detected them and obtained the glitch parameters, which are consistent with the results from \citet{2016ApJ...831...19L}.
 The remaining five glitches are new glitches discovered in this work, of which the second and fourth glitches are small glitches (with $\Delta\nu/\nu \sim 10^{-8}$ ) that may have been missed or mistaken for timing noise due to their small amplitudes. The other four glitches are relatively large, with their corresponding $\Delta\nu/\nu$ ranges from $10^{-7}$ to $10^{-6}$. The second and fourth glitches are detected between the previous glitches, which are detected about 467 days and 44 days following the first and third previous glitches and are not reported by previous work. The fractional glitch size of the second glitch (MJD $\sim$ 55419) is $\Delta\nu/\nu \sim 26(19)\times10^{-9}$, with a relative change in spin-down rate $\Delta \dot\nu /\dot\nu\sim -0.9(9)\times10^{-3}$.
The fourth glitch is detected on MJD 56080(2), with $\Delta\nu /\nu \sim 21.5(9)\times10^{-9}$ and $\Delta \dot\nu/\dot\nu \sim 0.6(6)\times 10^{-3}$, this is currently the smallest glitch we detected in this pulsar. The largest glitch we detected in this pulsar is the third one, which occurred on MJD 55463(2) with $\Delta\nu/\nu \sim 1.230(21) \times 10^{-6}$.
More detailed parameters of the rest glitches are given in Table~\ref{glitch solutions}. The pre- and post-glitch timing solutions of PSR J0007+7303 are shown in Table~\ref{pre and post glitch solutions}, we extract the values of frequency and its first and second derivatives to calculate the barking index, as shown in the last column.

\begin{figure}
	\includegraphics[width=1\columnwidth]{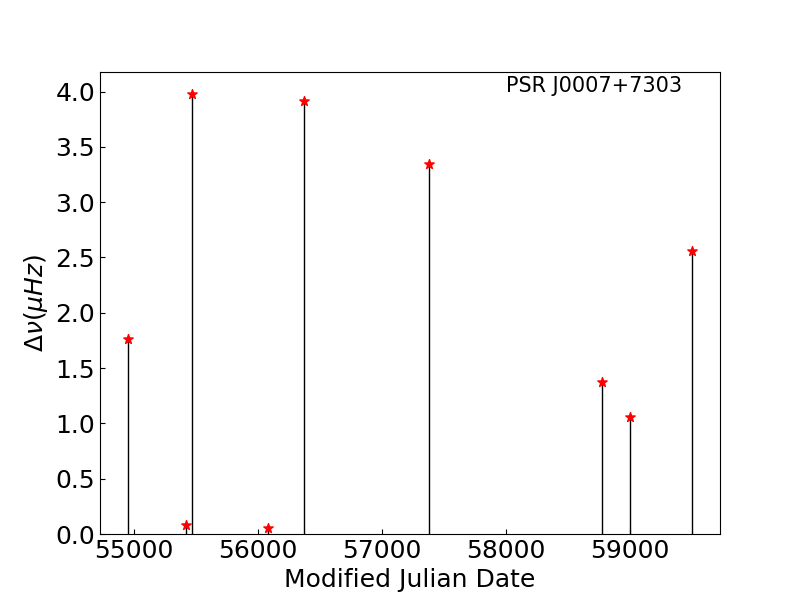}
    \caption{Nine glitches and their corresponding glitch epoch. The vertical coordinate is the spin frequency change of each glitch ($\Delta \nu$), and the horizontal axis represents the epoch of each glitch.}
    \label{watting time}
\end{figure}

\begin{figure}
	\includegraphics[width=1\columnwidth]{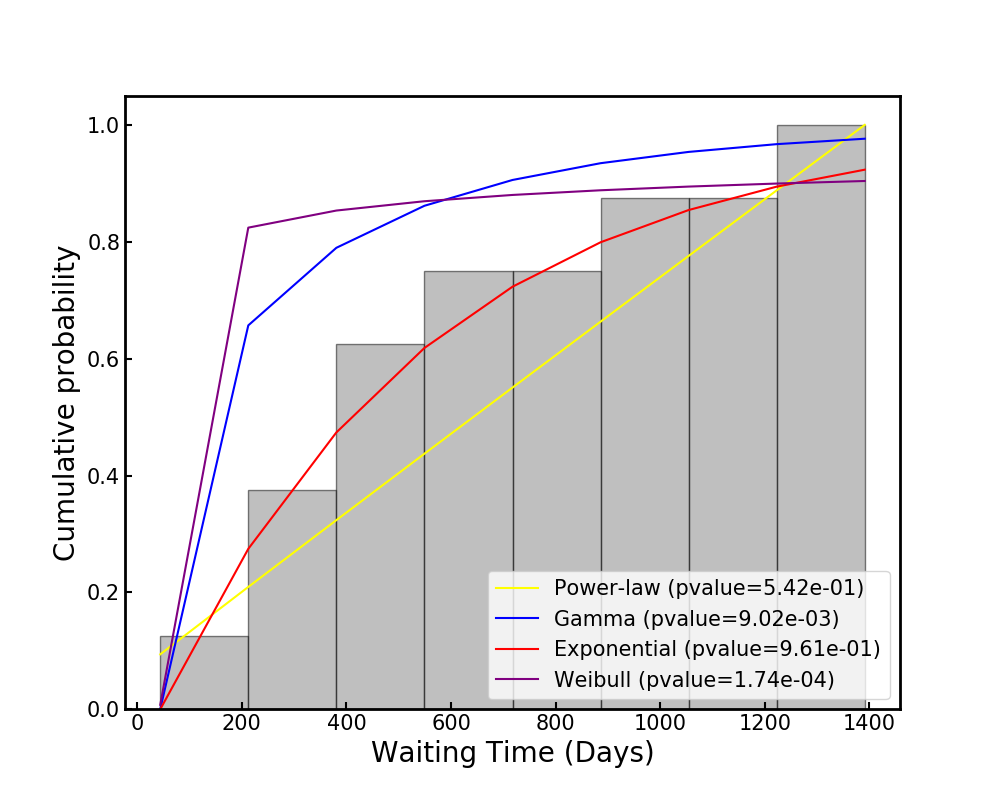}
    \caption{The cumulative distribution function of waiting time for nine glitches.}
    \label{cdf_v}
\end{figure}

Figure \ref{watting time} shows the temporal series of the glitch with time for the pulsar PSR J0007+7303, we choose the starting of observation as the starting point. 
 We statistically analyzed the nine glitches on PSR J0007+7303, and the magnitude of these glitches are shown in Fig \ref{watting time}. The vertical coordinate is the spin frequency change of each glitch ($\Delta \nu$), and the horizontal axis represents the epoch of each glitch. The interval between two glitches is called the waiting time, denoted by $T_{i}$. The interval $T_{2}$ between the second glitch and the third glitch is the shortest, only 44 days, and the third glitch has the largest size among all the detected ones, with small glitches detected before and after this glitch, respectively. The longest interval $T_{6}$ between the sixth glitch and the seventh glitch is about 1391 days. 
 We fitted the waiting times using the Exponential, Power-law, Weibull, and Gamma distributions, and found that the Exponential distribution provided the best fit with a statistical p-value of 0.961 (Fig \ref{cdf_v}). Different pulsars tend to show different glitch behaviors, for example, the glitch behavior of the Crab pulsar is irregular, while the glitch behavior of the Vela pulsar is relatively uniform, and the glitch events show relatively strong randomness. Although the pulsar is observed to show regular glitch behavior, with the increase in observation time, its glitch behavior may change. By analyzing the glitch waiting time of pulsars, we can conclude that the glitch behavior of pulsars is relatively random. It helps us conclude whether the glitch behavior of the pulsar is active or not.
\cite{Basu2022} reported the association of the glitch rate with the characteristic age and spin-down rate. The glitch rate is denoted as $R_{{\rm g}} = N/T$, where $T$ denotes the entire observation time and $N$ denotes the total number of glitches during the observation time $T$. The value of the glitch rate decreases with increasing characteristic age and increases with increasing spin-down rate. Glitch rate of this source $\sim$ 0.56 ${\rm yr}^{-1}$. Glitch occurs about once every two years in this pulsar, which is consistent with that the younger the age of the pulsar, the more likely it glitches. Quantitatively, according to Fig.10 in~\cite{2011MNRAS.414.1679E}, the statistic average glitching rate of a pulsar is 
\begin{equation}
\dot N_{\rm g}\simeq3\times 10^{-3}\times (\frac{|\dot\nu|}{10^{-15}~{\rm Hz~s^{-1}}})^{0.47}{\rm yr^{-1}}.
\label{glitch rate}
\end{equation}
For PSR J0007+7307, $\dot\nu=-3.61\times 10^{-12}~{\rm Hz~s^{-1}}$, 
therefore, $\dot N_{\rm g}\simeq 0.14~{\rm yr^{-1}}$, 
the glitch rate of PSR J0007+7307 is actually higher than prediction of Eq.(\ref{glitch rate}).


\subsection{Searching for the Variation of pulse profile and flux}

\begin{figure*}
\centering
\includegraphics[width=1\textwidth,height=10cm]{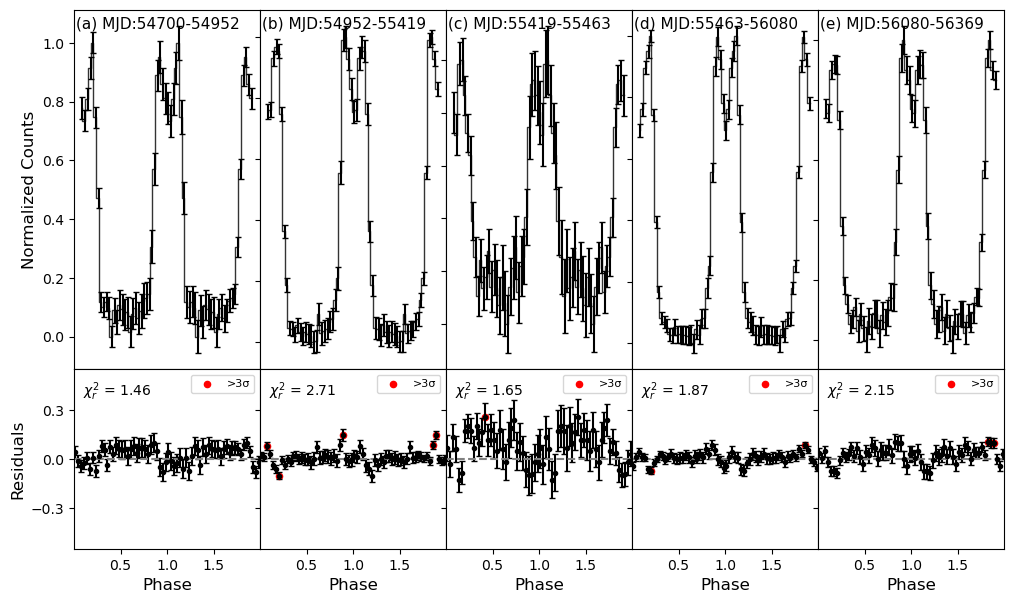}
\includegraphics[width=1\textwidth,height=10cm]{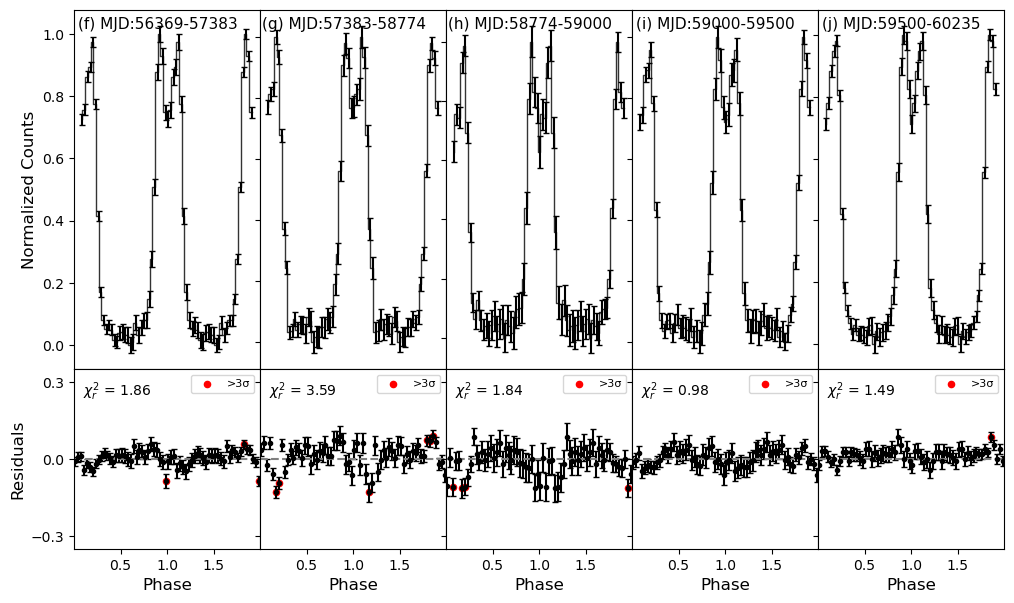}
\begin{center}
\caption{
Normalized gamma-ray pulse profile of PSR J0007+7303 over the corresponding time span. Two rotations are shown on the X-axis. Panels (a)$--$(j) represent post-glitch for glitches, respectively. The energy band is 0.1-10GeV.The lower panel shows the residuals between the profile of each panel and the total integral profile. The red circles indicate phase bins where the pulse profile residuals exceed the 3$\sigma$ threshold. Error bars represent $1\sigma$ Poisson uncertainties in the photon counts per bin.}
\label{progile_1-4}
\end{center}
\end{figure*}

\begin{figure}
	\includegraphics[width=1\columnwidth]{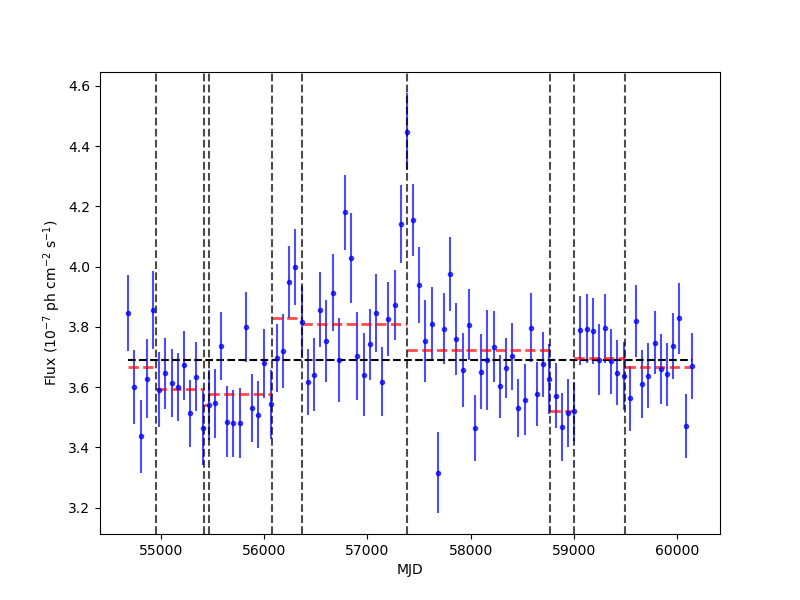}
    \caption{Gamama-ray flux evolution of PSR J0007+7303 from 2008 to 2023. The blue circles with error bars represent the flux measured in each 60-day time bin. The red dashed horizontal lines indicate the weighted mean flux within each time segment defined by known glitch epochs (black vertical dashed lines). The black dashed horizontal line shows the weighted mean flux over the entire dataset.}
    \label{flux}
\end{figure}

Searching for pulse profiles or flux variations associated with pulsar glitches is an important clue to determining their possible internal properties~\citep{Zhao2017,Kou2018,Wang2024} Therefore, we studied the changes in the pulse flux and pulse profile characteristics of pulsar PSR J0007+7307.
To study the effect of glitch events on the gamma-ray radiation, we divide the data span (August 4, 2008, to October 24, 2023) of this data into ten fragments using the glitch epoch. We analyze the effect of the glitch events on the gamma-ray radiation by observing variations of the pulse profiles before and after each glitch. Photons in the energy range of 150$-$30,000\ MeV are selected to obtain the integrated pulse profile, timing solutions, and pulse parameters used to assign photon phases are listed in Table~\ref{pre and post glitch solutions}. The exact phase of each photon is calculated by combining the photon data, telescope orbit data, and timing solutions, which can be realized by the Fermi-plugin of the {\sc Tempo2} software package.

To investigate whether the $\gamma-$ray pulse profile of the pulsar exhibits any significant temporal variations, we performed a segment-by-segment comparison of the normalized pulse profiles using a chi-square test. We divided the entire observational span into ten segments based on the glitch epoch, and compared each segment's profile to the integrated profile constructed from the full dataset.
Figure \ref{progile_1-4} shows the pulse profiles of each glitch interval. In each panel, displays the normalized $\gamma-$ray pulse profile for each glitch interval and residuals between total average pulse profile and profile of glitch intervals.
The pulse profile for each time interval exhibits a bimodal character, and we present the phases of the two cycles to facilitate the observation of changes in the pulse profiles.
The reduced chi-square($\chi^2_\mathrm{r}$) values range from 0.98 to 3.6, with the associated p-values spanning from 0.5 to below 0.0001. Notably, while some segments (e.g., segments (f) and (g)) show relatively large $\chi^2_\mathrm{r}$ values and small p-values indicating statistically significant differences, the number of bins deviating beyond the 3$\sigma$ level remains small. 
In most segments, over $95\%$ of the phase bins remain within the 3$\sigma$ residual range, suggesting that the profile morphology is largely stable. 
For instance, segment (a) (MJD $54700-54952$) and segment (i) (MJD $59000–59500$) have all 32 bins within 3$\sigma$. These results suggest that, although minor fluctuations in the pulse shape may exist, the overall $\gamma-$ray pulse profile remains statistically consistent throughout the observation span.

To check the long-term stability of the flux, we perform a flux analysis of this source using the Fermi-LAT maximum-likelihood Science Tool ``gtlike". We select data with an energy range of 0.1$-$300\ GeV and consider a $10^{\circ}$ radius region of interest (ROI) centered at the J0007$+$7303, events from the point source or Galactic diffuse class (event class = 128) and data from both the front and back sections of the tracker (evttype = 3), and also select photon events with zenith angles below $90^{\circ}$ to avoid contamination from the Earth's atmosphere.
We construct background emission model, which include all the catalog sources of the Fermi LAT Fourth Source Catalog (4FGL-DR4, \citep{ballet2024fermilargeareatelescope}) within $10^{\circ}$ of the center of the ROI, the galactic diffuse emission (gll\_iem\_v07) and the isotropic diffuse emission (iso\_P8R3\_SOURCE\_V3\_v1)
obtained from the Fermi Science Support Center.
We add 10 degrees to PSR J0007+7303’s ROI to account for sources that lie outside our data region. We also use a limit of five degrees from the center of the search region to reduce the number of sources with free parameters. A power-law with an exponential cutoff model is used for our source.
Finally, we fit each bin (60 days) of the flux evolution through the binned likelihood analysis (gtlike). 
The flux evolution of PSR J0007+7303 is shown in Figure~\ref{flux}, where each blue point with error bar represents the observed $\gamma-$ray flux in a 60-day bin. To further investigate whether the $\gamma-$ray flux variations are associated with glitch events, we divided the light curve into segments defined by the known glitch epochs and performed constant flux fitting within each segment. The best-fit flux in each segment was obtained through weighted averaging, taking into account the flux uncertainties. We then evaluated the significance of flux differences across glitch boundaries. In Figure~\ref{flux}, the red dashed lines show the weighted mean flux within each segment defined by glitch epochs, while the horizontal black dashed line represents the weighted mean flux across the entire data span. Vertical dashed lines mark the epochs of detected glitches. To obtain a robust conclusion, we calculated the weighted mean flux differences across each glitch and evaluated their significance by comparing the flux difference to its associated uncertainty. Among all glitch events, only the increase in the weighted mean flux after glitch 4 (MJD 56080) appears to be notable, with a significance level of approximately 3.9$\sigma$. For all other glitches, the significance of flux changes remains below the 3$\sigma$ threshold. However, the post-glitch-4 segment contains only five data points, which weakens the statistical reliability of this result. Therefore, we do not consider it as strong evidence for glitch-induced flux variation. Overall, we find no clear or statistically significant $\gamma-$ray flux changes associated with glitches in PSR J0007+7303.

\subsection{Phase Average Spectral variability Analysis}
To detect potential spectral alterations in PSR J0007+7303 in the vicinity of glitch events, we segment the Fermi-LAT data into ten distinct intervals centered on the glitches. Subsequently, we employed the power law model with an exponential cutoff to carry out a likelihood analysis for the pulsar across all ten of these time bins, examining spectral variability before and after the glitches.
For our average spectral analysis, we build background emission model to perform the likelihood
analysis including gamma-ray sources within $10^{\circ}$ of PSR J0007+7303 included in the 4FGL-DR4, isotropic and Galactic diffuse emission models (``gll\_iem\_v07.fits" and ``iso\_P8R3\_SOURCE\_v1.txt"). And we also have add 10 degrees to PSR J0007+7303’s ROI to account for sources that lie outside our data region. Five degrees are the spectral parameters of background sources set to free. Additionally, we fixed the spectral parameters of SNR CTA-1 and S5 0016+73.

\begin{figure}
	\includegraphics[width=1\columnwidth]{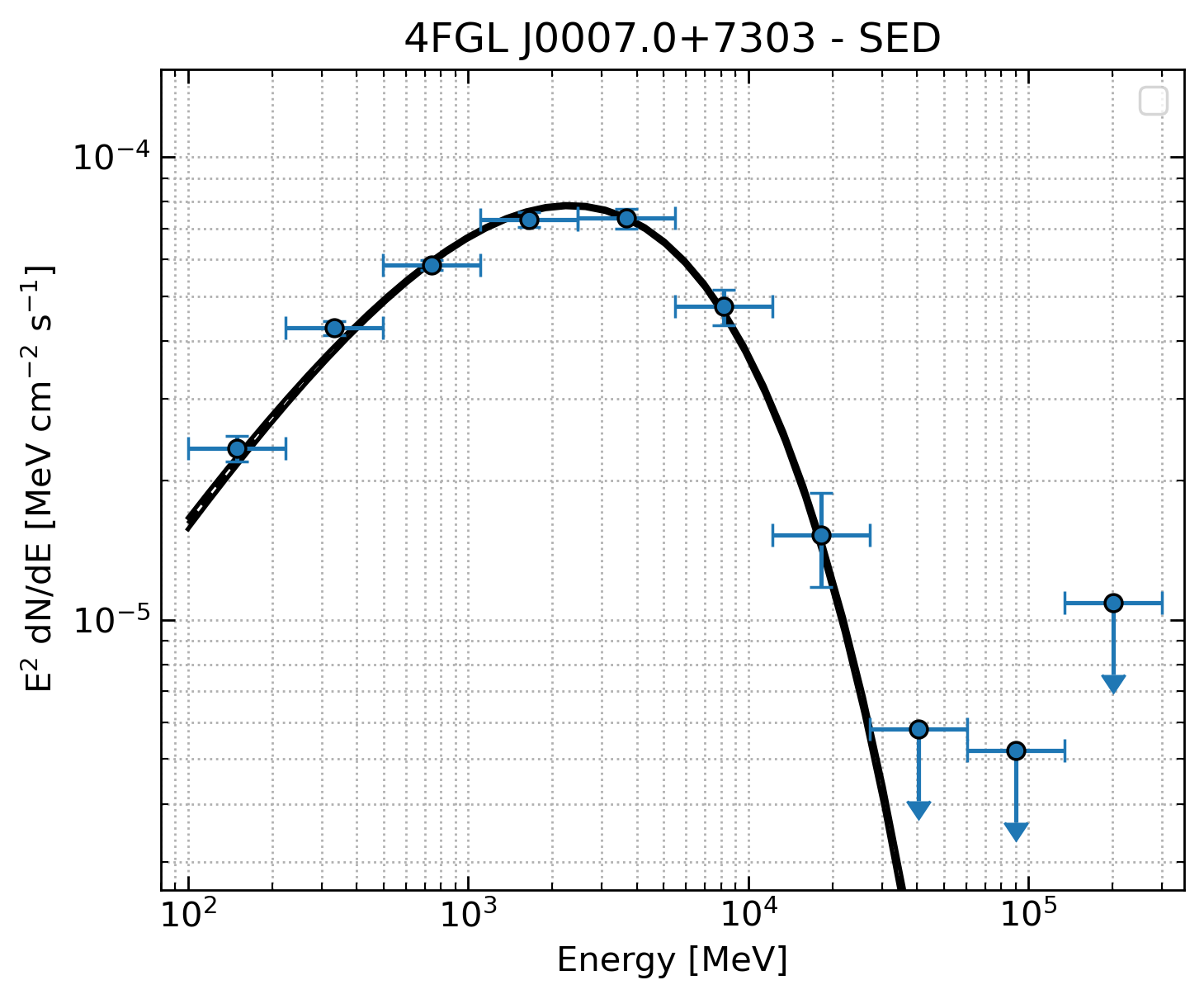}
    \caption{Spectra of PSR J0007+7303 during time-span before the first glitch. Solid line presents the best-fit model according to the `PLSuperExpCutoff' model.}
    \label{SED}
\end{figure}

The above analysis was implemented using the easyFermi software\citep{2022A&C....4000609D}. 
The spectrum of PSR J0007+7303 can be well fitted with a power-law with an exponential cutoff, which can be expressed as: 

 \begin{equation}
   \dfrac{{\rm d}N}{{\rm d}E} =N_{0} \left(\frac{E}{E_{0}} \right)
   ^{\Gamma}{\rm exp}\left[-\left(\frac{E}{E_{C}} \right)^b\right],
\label{equation:PLEC model}
 \end{equation}
 where $N$ is the number of photons per unit time,
 unit area and unit energy; $E$ is the energy of photons; $N_{0}$ is the normalization constant; $E_{0}$ is the scale factor of energy; $\Gamma$ is the spectral power-law index; and $E_{C}$ is the cutoff energy. 
Figure \ref{SED} presents the spectra before the first glitch fitted by fermipy~\citep{Wood2017}. Spectral parameters of PSR J0007+7303 are present in Table \ref{pre and post SED parameters }. The test statistic (TS) was employed to evaluate the
significance of the gamma-ray fluxes coming from the sources. \citet{2016ApJ...831...19L} has reported the definition of TS. The b index were left free and the evolution of spectral parameters were represented in Figure \ref{spectral_evolution}. We have not found a significant variability on each parameter of spectrum.

\begin{table*}
\centering
\caption{Parameters of spectra of PSR J0007+7303 at different inter-glitch intervals.}

\label{pre and post SED parameters }
\begin{center}
    \renewcommand{\arraystretch}{1.4}
    \setlength{\tabcolsep}{3pt}      
\resizebox{1\textwidth}{!}{
\begin{tabular}{ccccccccc}
  \hline
    \hline       
Time Span  & Spectral index  & Cutoff Energy  & b   & FLux   & \\
   (MJD)        &                 & (GeV)      &     & ($10^{-7} ph\ cm^{-2} \ s^{-1}$)       &  \\
    \hline 
54700 -- 54952   &1.47(8)  &5.39(2)   & 1   &3.51(6)   \\     
54952 -- 55419   &1.43(6)   &4.31(1)  &1     &3.55(4) \\ 
55419 -- 55463   &1.53(16)  &5.73(134)  &1    &3.54(14) &\\ 
55463 -- 56080   &1.52(4)  &5.86(2)   &1    &3.59(4)  & \\    
56080 -- 56369   &1.49(4)  &5.49(12)  &1    &3.75(5)  & \\   
56369 -- 57385   &1.54(2)  &5.98(27)  &1      &3.81(3)  & \\  
57385 -- 58774   &1.50(13)  &5.58(19)  &1        &3.74(2) &  \\ 
58774 -- 59000   &1.47(16)   &5.12(4)   &1       &3.50(5)  & \\  
59000 -- 59500   &1.48(5)     &5.39(2)   &1       & 3.61(4)  & \\  
59500 -- 60235   &1.47(12)     &5.39(20)   &1      &3.61(3)   & \\        
\hline
\end{tabular}}
\end{center} \vspace{-0.4cm}
\end{table*}

\begin{figure}
	\includegraphics[width=1\columnwidth]{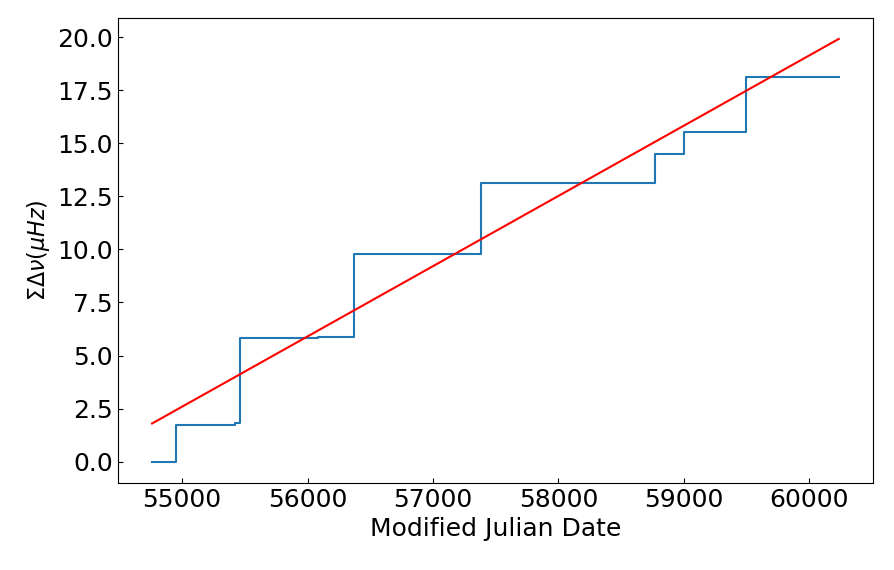}
    \caption{Cumulative change of pulsar spin frequency at glitches over time. The straight line is the least-squares fit, with $\dot\nu_{\rm g}$ as the slope.}
    \label{cumSize}
\end{figure}


\begin{figure}
\centering
	\includegraphics[width=0.8\columnwidth]{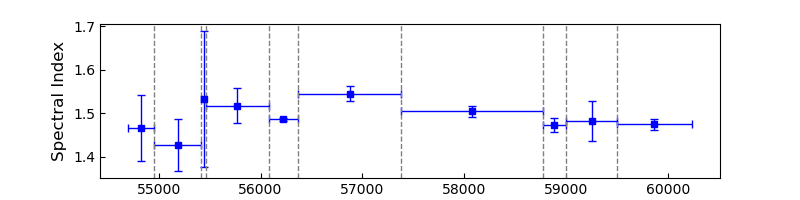}\\
	\includegraphics[width=0.8\columnwidth]{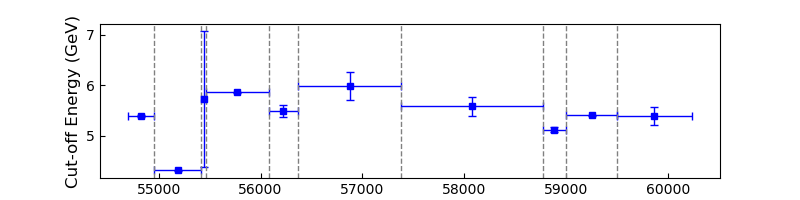}\\
	
    \caption{From top to bottom: cutoff energy of PSR J0007+7303 in the range 0.1–300 GeV for epochs separated by the glitches, the values of spectral index in the model of a power law with exponential cutoff; The dashed vertical gray lines mark the times of glitches.}
    
    \label{spectral_evolution}
\end{figure}

\section{Discussion}\label{sec:Discussion}
We performed a timing analysis of PSR J0007+7303 and detected a total number of nine glitches, five of which are newly discovered in this study. 
Seven out of the nine glitches are large ones with fractional size $\sim 10^{-6}$, while only the second and the fourth ones are small glitches with fractional size $\sim 10^{-8}$.
We have compared the pre-glitch and post-glitch pulse profiles, gamma-ray flux, and spectral fitting parameters of all the nine glitches, no significant variations have been detected, indicating an internal origin of these glitches. 
The non-detection of radiative changes is consistent with observations of other young pulsars, such as the Crab pulsar~\citep{2017A&A...597L...9V,2018MNRAS.478.3832S,2022ApJ...932...11Z}, although they are observed in different energy bands.
Besides, we have generated the cumulative distribution function of glitch waiting time (Fig \ref{cdf_v}), we found that a power law function is preferred.
The cumulative change of pulsar spin frequency at glitches increases steadily with time, as shown in Fig \ref{cumSize}.

With the discovery of five new glitches form PSR J0007+7303, we can calculate its long-term glitch activity.
For one single pulsar, the glitch activity is defined as~\citep{2000MNRAS.315..534L,2011MNRAS.414.1679E,2017A&A...608A.131F}
\begin{equation}
\dot\nu_{\rm g}=\frac{\sum_{i}\Delta\nu_{i}}{\Delta T},
\end{equation}
where $\Delta T$ is the total observation time over which the pulsar has been searched for glitches, $\Delta\nu_{i}$ is the step increase in frequency for the glitch $i$ of this pulsar.
Assuming PSR J0007+7303 has only experienced the nine glitches as shown in Table \ref{glitch solutions}, we have $\Delta T=15~{\rm yrs}$ and $\dot\nu_{\rm g}=3.83\times 10^{-14}~{\rm Hz/s}$.
Therefore, 
\begin{equation}
\label{glitch activity}
\dot\nu_{\rm g}/|\dot\nu|=1.06\times 10^{-2},
\end{equation}
this is consistent with the most recent result $\dot\nu_{\rm g}/|\dot\nu|=0.01\pm 0.001$ in \cite{2017A&A...608A.131F}.
Theoretically, if the glitches originate from neutron star interior, $\dot\nu_{\rm g}/|\dot\nu|$ reflects the fractional moment of inertia of neutron star crustal superfluid.
Eq.(\ref{glitch activity}) means that the moment of inertia of crustal superfluid get involved in glitches amounts to $1.06\%$ of this pulsar's total moment of inertia if the non-dissipative entrainment effect due to Bragg scattering~\citep{2005NuPhA.759..441C,2005NuPhA.748..675C,2006IJMPD..15..777C,2005NuPhA.747..109C,2006NuPhA.773..263C,2012PhRvC..85c5801C} is not considered, and the consistency with the result of \cite{2017A&A...608A.131F} means the superfluid angular momentum reservoir of PSR J0007+7303 is similar with that of other glitching pulsars, which further supports the internal origin of glitches of PSR J0007+7303.
Note that, if the surface thermal temperature of PSR J0007+7303 could be measured accurately, Eq.(\ref{glitch activity}) could be used to constrain the equation of state and mass of this pulsar~\citep{2015SciA....1E0578H}.

Although PSR J0007+7303 shows many similarities with other glitching pulsars as we discussed above, its post-glitch recoveries are different from standard glitches.
Vela pulsar glitches are viewed as the standard glitches, and the post-glitch recoveries of Vela pulsar glitches have been described successfully within the vortex creep model~\citep{1993ApJ...409..345A}.
Typically, Vela pulsar glitches include several rapid exponential components and long-term linear increases in spin down rate~\citep{1990Natur.345..416F,2018IAUS..337..197M}. If the recovery process will not be interrupted by the following one glitch, the rotational spin down rate will gradually return to the value predicted by linear fitting to the pre-glitch data points, the rotational frequency will also recover partly.
However, the post-glitch behaviors of glitches in PSR J0007+7303 are different from that of Vela pulsar, in spite of its young and similar characteristic age with the Vela pulsar.
As we can see from Fig.\ref{all_glitch}, glitches of PSR J0007+7303 show no apparent exponential or linear recoveries, most of its glitches appear as simple steps in frequency that do not recover at all.
The fifth and sixth glitches show recoveries to some extend (middle panels of Fig.\ref{all_glitch}), but the spin down rate shows no apparent trends towards its pre-glitch value.

How to understand the no recovery behaviors of glitches in the young pulsar J0007+7303?

From the perspective of the pure starquake model, it can not account for all the observations of PSR J0007+7303 consistently. Firstly, three glitches of fractional sizes $\sim 10^{-6}$ occurred in PSR J0007+7303 within about $1900~{\rm days}$, as shown in Table \ref{glitch solutions}. However, theoretical calculations have shown that, the pure starquake model is insufficient to result in glitches of fractional size of $\Delta\nu/\nu\sim 10^{-6}$ in human life timescale~\citep{1971AnPhy..66..816B}, recent developments in starquake model have supported this conclusion~\citep{2003ApJ...588..975C,2008A&A...491..489Z, 2020MNRAS.491.1064G}. Secondly, starquake could induce changes in pulse profile and/or flux by affecting the distribution of magnetic fields, these were not consistent with observations of PSR J0007+7303 in the gamma-ray band. Thirdly, the post-glitch relaxation theory in the neutron star starquake model has not been constructed, so the presence of no recovery in glitches of PSR J0007+7303 is hard to understand in the starquake model. 
Note that, even if the starquake merely serves as the trigger and vortex unpinning contributes to the glitch respectively~\citep{2018MNRAS.473..621A}, the question of why no recovery process was observed can not be avoided and should be interpreted according to the superfluid vortex model due to the third point as stated above.

From the perspective of the superfluid model, several explanations have been proposed for the no recovery phenomenon.

\citet{2014MNRAS.438L..16H} have shown that, the vortex unpinning paradigm can explain different kinds of relaxation.
They claimed that, if the glitch takes place in a strongly pinned region, glitch appears as a step in frequency and shows no recovery.
Apart from this, G{\"u}gercino{\u{g}}lu have considered the post-glitch relaxation of radio pulsars and magnetars in terms of vortex creep across flux tube. 
The corresponding relaxation timescale is given by~\citep{Gugercinoglu2017}
\begin{equation}
\label{timescale}
\tau=60\left(\frac{|\dot\Omega|}{10^{-10}~{\rm rad~s^{-2}}}\right)^{-1}\left(\frac{T_{\rm in}}{10^8~{\rm K}}\right)\left(\frac{R}{10^6~{\rm cm}}\right)^{-1}{x_{\rm p}}^{1/2}\left(\frac{m_{\rm p}^{*}}{m_{\rm p}}\right)^{-1/2}\left(\frac{\rho}{10^{14}~{\rm g~cm^{-3}}}\right)^{-1/2}\left(\frac{B_{\phi}}{10^{14}~{\rm G}}\right)^{1/2}~{\rm day},
\end{equation}
where $\dot\Omega=-2.2682\times 10^{-11}~{\rm rad~s^{-2}}$, $T_{\rm in}$ is the internal temperature, $R$ is the location of toroidal magnetic field region, $x_{\rm p}$ is the fraction of protons in neutron star core, $m_{\rm p} (m_{\rm p}^{*})$ is the bare (effective) mass of protons, $\rho$ is the density where glitch occurs, and $B_{\phi}$ is the internal toroidal magnetic field strength.
For PSR J0007+7303, the surface magnetic field is $B_{\rm d}=1.08\times 10^{13}~{\rm G}$~\footnote{Data taken from the ATNF Pulsar Catalogue version 2.5.1 at the url: https://www.atnf.csiro.au/research/pulsar/psrcat/.}, smaller than that of the high magnetic field pulsar J1119-6127, so $B_{\phi}=3.3\times 10^{14}~\rm{G}$ according to Eq.(6) in \cite{Gugercinoglu2017}.
For the internal temperature $T_{\rm in}$, if we use Eq.(15) in \cite{Gugercinoglu2017} or equivalently $T_{\rm in}=3.3\times 10^{8}(\tau_{\rm c}/1~{\rm yr})^{-1/6}~{\rm K}$ based on modified Urca cooling law~\citep{2009PhRvL.102n1101G}, it gives $T_{\rm in}=1.685\times 10^{8}~{\rm K}$.
However, \citet{2010ApJ...725L...6C} has limited its surface temperature of a $10~{\rm km}$ neutron star to be between $4.8\times 10^5~{\rm K}$ and $5.3\times 10^5~{\rm K}$ using blackbody fitting to the X-ray spectrum.
If this blackbody temperature is reliable, translating this surface temperature into internal temperature through the prescription of \cite{1983ApJ...272..286G} gives $T_{\rm in}\sim 2.757\times 10^{7}~{\rm K}$. 
As of parameters $R$, $m_{\rm p}^{*}/m_{\rm p}$ and $\rho$, they depends on neutron star equation of state.
Using parameters of model 1, model 2, model 3 in \cite{Gugercinoglu2017} and $T_{\rm in}=1.685\times 10^{8}~{\rm K}$, the relaxation timescale will be $\tau\simeq 239~{\rm days}, 198~{\rm days}$, or $160~{\rm days}$.
While if $T_{\rm in}\sim 2.757\times 10^{7}~{\rm K}$ is adopted, the relaxation timescale will be $\tau \simeq 39~{\rm days}, 32~{\rm days}$, or $26~{\rm days}$ respectively.
The absence of post-glitch recoveries of PSR J0007+7303 means a much long relaxation timescale, therefore, confronting the theoretical timescales with post-glitch observations of PSR J0007+7303, an equation of state that is stiffer than model 1 in \cite{Gugercinoglu2017} is more preferred, given the temperature uncertainty of PSR J0007+7303.

\section{Summary}\label{sec:Summary}
Using the Fermi-LAT telescope, we perform a timing analysis of PSR J0007+7303 based on 15 years of observation. Following is our main conclusion.
\begin{enumerate}
\item We detect a total of nine glitches, five of which are newly detected, while four glitches have already been reported, we update the parameters for these three previously reported glitches. The magnitude of the glitches detected in this work ranges from $10^{-6}-10^{-9}$. Two of the nine glitches are small glitches. There is no significant exponential recovery following any of the nine glitches.
\item We calculate the waiting time and analyze the glitch rate for this source. We find that the second glitch (MJD $\sim$55419) is followed by a large glitch (MJD $\sim$55463) only 44 days later. The glitch rate is approximately $\sim 0.56~{\rm yr}^{-1}$ which is consistent with the young age of the star. The waiting time of glitches on the pulsar is random, with the longest interval exceeding three years and the shortest being only 44 days.
\item We have analyzed the effect of glitches on the evolution of the flux, pulse profiles and average phase spectral. No significant variation were detected. 

\end{enumerate}

Our results indicate an internal origin of these glitches in PSR J0007+7303. We look forward to conducting long-term pulsar timing monitoring of this pulsar in the future, in order to collect more glitch events, which will help us to better understand the physical mechanisms behind glitches.

\section*{Acknowledgements}

     This work is supported by the open research project funded by the Key Laboratory of Xinjiang Uyghur Autonomous Region (2022D04015), the Major Science and Technology Program of Xinjiang Uygur Autonomous Region (No.2022A03013-4, 2022A03013-2), the Guizhou Province Science and Technology Foundation (No. ZK[2022]304), the Guizhou Provincial Science and Technology Projects (Nos. QKHFQ[2023]003, QKHPTRC-ZDSYS[2023]003, QKHFQ[2024]001-1), the Scientific Research Project of the Guizhou Provincial Education (Nos. KY[2022]132, KY[2022]137, KY[2022]123), the Foundation of Education Bureau of Guizhou Province, China (Grant No. KY (2020) 003), the National Natural Science Foundation of China (No. 12041304), Zhejiang Provincial Natural Science Foundation of China (grant No. LQ24A030002), the Project funded by China Postdoctoral Science Foundation No. 2023M743517, science and technology department of Gansu province (No. 20JR5RA481), the National Natural Science Foundation of China (NSFC) project (Nos. 12403046), Natural Science Basic Research Program of Shaanxi (Program No. 2024JC-YBQN-0036), the Natural Science Foundation of Xinjiang Uygur Autonomous Region (No. 2023D01E20), Sichuan Provincial Natural Science Foundation Project No.2025ZNSFSC0878, the Tianshan talents program, the Academic New Seedling Fund Project of Guizhou Normal University (No.[2022]B18), the Natural Science Foundation of China ($12203093$), the National Key Research and Development Program ($2022YFA1603104$), the Major Science and Technology Program of Xinjiang Uygur Autonomous Region ($2022A03013-2$). We acknowledge the use of Fermi-LAT public data.

\bibliography{sample631}
\bibliographystyle{aasjournal} 

\end{document}